\documentclass{elsart}
\usepackage{graphicx}

\begin{document}

\begin{frontmatter}

\title{Energy Spectra of Elemental Groups of Cosmic Rays: 
Update on the KASCADE Unfolding Analysis}

\author[fzk]{W.D. Apel},
\author[uni]{J.C. Arteaga\thanksref{r0}},
\author[fzk]{A.F. Badea}, 
\author[fzk]{K. Bekk},
\author[fzk,uni]{J. Bl\"umer},
\author[fzk]{H. Bozdog},
\author[rum]{I.M. Brancus}, 
\author[sie]{M. Br\"uggemann}, 
\author[sie]{P. Buchholz}, 
\author[uni]{F. Cossavella},
\author[fzk]{K. Daumiller},
\author[uni]{V. de Souza},
\author[fzk]{P. Doll},
\author[fzk]{R. Engel}, 
\author[fzk]{J. Engler}, 
\author[uni]{M. Finger\thanksref{corr}}, 
\author[wup]{D. Fuhrmann}, 
\author[fzk]{H.J. Gils},
\author[wup]{R. Glasstetter}, 
\author[sie]{C. Grupen}, 
\author[fzk]{A. Haungs},
\author[fzk]{D. Heck},
\author[uni]{J.R. H\"orandel\thanksref{now}}, 
\author[fzk]{T. Huege},
\author[fzk]{P.G. Isar},
\author[wup]{K.-H. Kampert}, 
\author[uni]{D. Kang}, 
\author[sie]{D. Kickelbick}, 
\author[fzk]{H.O. Klages},
\author[sie]{Y. Kolotaev}, 
\author[pol]{P. {\L}uczak},
\author[fzk]{H.J. Mathes}, 
\author[fzk]{H.J. Mayer},
\author[fzk]{J. Milke},
\author[rum]{B. Mitrica},
\author[fzk]{S. Nehls},
\author[fzk]{J. Oehlschl\"ager}, 
\author[fzk]{S. Ostapchenko}, 
\author[sie]{S. Over},
\author[rum]{M. Petcu},
\author[fzk]{T. Pierog},
\author[fzk]{H. Rebel},
\author[fzk]{M. Roth},
\author[fzk]{G. Schatz}, 
\author[fzk]{H. Schieler}, 
\author[fzk]{F. Schr\"oder}, 
\author[buc]{O. Sima}, 
\author[uni]{M. St\"umpert},
\author[rum]{G. Toma},
\author[fzk]{H. Ulrich},
\author[fzk]{J. van Buren},
\author[sie]{W. Walkowiak},
\author[fzk]{A. Weindl},
\author[fzk]{J. Wochele}, 
\author[fzk]{M. Wommer}, 
\author[pol]{J. Zabierowski}

\address[fzk]{Institut f\"ur Kernphysik, Forschungszentrum Karlsruhe, Germany}
\address[uni]{Institut f\"ur Experimentelle Kernphysik, Universit\"at 
Karlsruhe, Germany}
\address[rum]{National Institute of Physics and Nuclear Engineering, 
Bucharest, Romania}
\address[sie]{Fachbereich Physik, Universit\"at Siegen, Germany}
\address[wup]{Fachbereich Physik, Universit\"at Wuppertal, Germany}
\address[pol]{Soltan Institute for Nuclear Studies, Lodz, Poland}
\address[buc]{Department of Physics, University of Bucharest, Romania}

\thanks[corr]{corresponding author, {\it E-mail address:} finger@ik.fzk.de}
\thanks[r0]{now at: Universidad Michoacana, Morelia, Mexico}
\thanks[now]{now at: Radboud University Nijmegen, The Netherlands}

\begin{abstract}
The KASCADE experiment measures extensive air showers induced by cosmic rays in the
energy range around the so-called knee. The data of KASCADE have
been used in a composition analysis showing the knee at 3--5~PeV to be caused by a steepening
in the light-element spectra~\cite{Ant05}. 
Since the applied unfolding analysis depends crucially on simulations of air showers, 
different high energy hadronic interaction models (QGSJet and SIBYLL) were used. 
The results have shown a strong dependence of the relative abundance of the 
individual mass groups on the underlying model.
In this update of the analysis we apply the unfolding method with a different 
low energy interaction model (FLUKA instead of GHEISHA) in the simulations.  
While the resulting individual mass group spectra do not change significantly, 
the overall description of the measured data improves by using the FLUKA model. 
In addition data in a larger range of zenith angle are analysed.
The new results are completely consistent, i.e. there is no 
hint to any severe problem in applying the unfolding analysis method to KASCADE data.
\end{abstract}

\end{frontmatter}

\section{Introduction}

Due to the rapidly falling intensity with increasing energy, 
cosmic rays of energies above $10^{15}\,$eV can be studied only 
indirectly by observations of extensive air showers (EAS)
which are produced by the interactions of cosmic 
particles with nuclei of the Earth's atmosphere.  
The observation of a change of the power law slope~\cite{knee} 
of the size spectrum of EAS and consequently of the all-particle 
energy spectrum at $\sim 3\cdot 10^{15}\,$eV 50 years ago 
has not yet been convincingly explained~\cite{Hau03}. 
Several theories for the origin of the knee predict different
knee positions for particles of different primary mass.
Therefore the energy spectra of single elements or at least
mass groups are of considerable interest.
Indeed, recent analyses \cite{Agl04,Ant05} find a steepening in 
the energy spectra of the light components in the knee region.
Whereas the measurement method of detecting air showers
alleviates statistical problems, one has to rely on the results of
simulations and the description of hadronic interactions while 
reconstructing the properties of the primary particles. 
Since the required energy and important kinematic regions of these 
interactions are beyond the range of collider or fixed target 
experiments, the interaction models used are uncertain and differ in 
their predictions. On the other hand, a thorough analysis of EAS 
data offers the opportunity of testing~\cite{Mil06} and 
improving the validity of these hadronic interaction models.

The KASCADE-Grande experiment~\cite{Nav04}, located on site of the Forschungszentrum
Karlsruhe (Germany), is designed to measure EAS in the energy range between
0.5~PeV and 1~EeV. The installation consists of the original KASCADE~\cite{Ant03} experiment
and an extension by the Grande array, covering an effective area of 0.5~km$^{2}$. 

The data analysis pursued by the KASCADE collaboration invokes 
an unfolding procedure of the two-dimensional shower size 
spectrum (total electron number vs.~muon number) into energy spectra 
of five individual mass groups~\cite{Ant05}. 
Despite the success of this method for the reconstruction 
of the shape of the spectral forms, a strong dependence of the result for the 
elemental abundances on the interaction model underlying the analysis was 
found. Also, an insufficient description of the measured data by the
employed simulations has been demonstrated.
While in~\cite{Ant05} the results for the analysis based on two different 
high energy interaction models were compared, we give
in the present paper an update of the composition analysis concentrating on 
the influence of the low energy interaction model on the result.
Furthermore, the analysis is repeated for measured data of different zenith angle 
intervals, thus testing the consistency of the analysis. 

\section{Composition analysis of the KASCADE shower size spectrum}

\begin{figure}[ht]
\includegraphics*[width=0.5\textwidth]{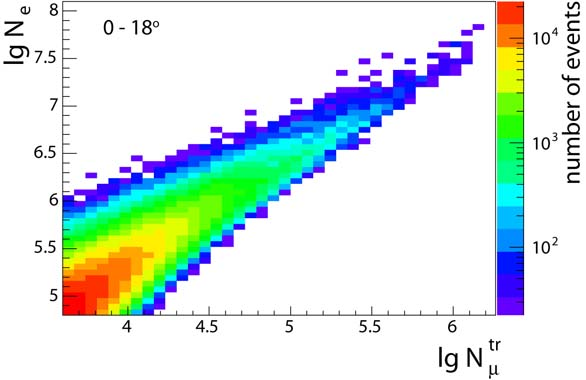}
\includegraphics*[width=0.5\textwidth]{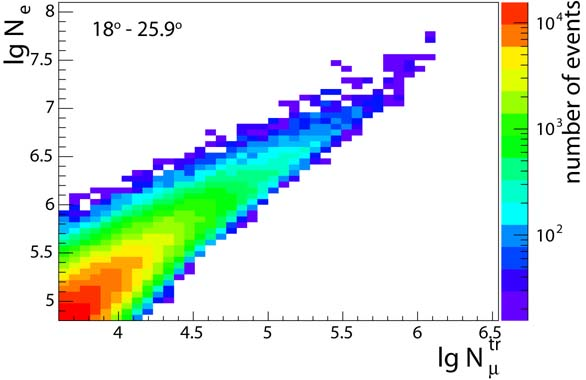}
\includegraphics*[width=0.5\textwidth]{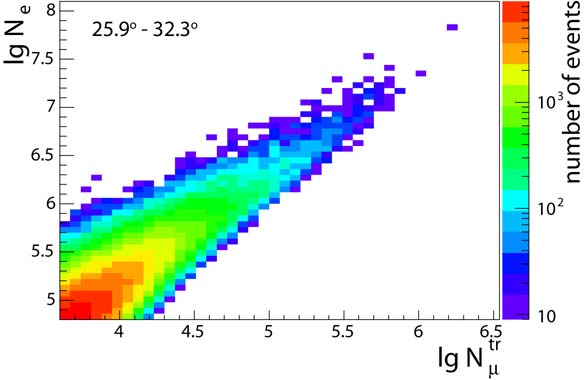}
\caption{Two-dimensional shower size spectra as measured by KASCADE for three different 
zenith angular ranges, where the first one shows similar data as used for 
the analysis described in~\cite{Ant05}. The effective time of the measurements amounts 
to 840~days.}
\label{2dimspec}
\end{figure}

Starting point of the analysis is the so-called two-dimensional shower size spectrum, i.e.
the number of measured EAS depending on the electron number $\lg N_{e}$ and the truncated
muon number $\lg N_{\mu}^{tr}$ (number of muons with shower core distances 
between $40\,$m and $200\,$m).
In Fig.~\ref{2dimspec} this spectrum is given for showers inside the KASCADE array for three
different ranges of inclination angle: EAS between $0^\circ$ and $18^\circ$, between 
18$^{\circ}$ and 25.9$^{\circ}$, and between 25.9$^{\circ}$ and 32.3$^{\circ}$, respectively.

The content $N_{j}$ of each histogram cell $j$ can be written as
\begin{equation}
N_{j} = C \sum_{A=1}^{N_{A}}
\int_{-\infty}^{+\infty} \frac{dJ_{A}}{d\lg E} \/
\,p_{A}
\,d\lg E .
\label{integral1}
\end{equation}
$C$ is a normalizing constant (time, aperture), and the sum is carried out over
all primary particle masses $A$. The functions $p_{A} = p_{A}(\lg N_{e,j}, \lg
N_{\mu,j}^{tr} | \lg E)$ give the probability for an EAS with primary energy $E$ and mass
$A$ to be measured and reconstructed with shower sizes $N_{e,j}$ and $N_{\mu,j}^{tr}$.
The probabilities $p_{A}$ include shower fluctuations, efficiencies, and reconstruction
resolution. For reasons of clarity integration over solid angle and cell
area is omitted in Eqn.~\ref{integral1}.
In case of KASCADE data $p_{A}$ is dominated by the shower fluctuations, whereas
reconstruction systematics play an inferior role~\cite{Ant05}. 
The data range was chosen in a way to minimize influences from inefficiencies.

Adopting this notation the two-dimensional shower size spectrum is regarded as a set of
coupled integral equations. In the analysis the primary particles H, He, C, Si, and Fe are
chosen as representatives for five mass groups. The corresponding probabilities $p_{A}$
are determined by Monte Carlo simulations using CORSIKA~\cite{Hec98} and a detailed
GEANT~\cite{GEA} based simulation of the experiment. To solve the equation system for 
the mass group energy spectra, the unfolding algorithm proposed by Gold~\cite{Gol64} is 
applied. Details of the selection, reconstruction, and the analysis can be 
found in Ref.~\cite{Ant05}. In particular, in that paper the determination and definition 
of the systematic uncertainties and inaccuracies of the analysis chain as well as the 
applicability of the chosen unfolding algorithm method is discussed in detail. 

\section{Influence of low energy hadronic interaction model}

In the original analysis~\cite{Ant05} the probabilities $p_{A}$ were determined using
the high energy hadronic interaction models QGSJet~\cite{Kal93} (2001 version) and
SIBYLL~\cite{Eng99} (version 2.1) in the simulations. For both cases the 
GHEISHA~\cite{Fes85} code was used for interactions with laboratory energy of $<80$~GeV. 
\begin{figure}[ht]
\begin{center}
\includegraphics*[width=0.9\textwidth]{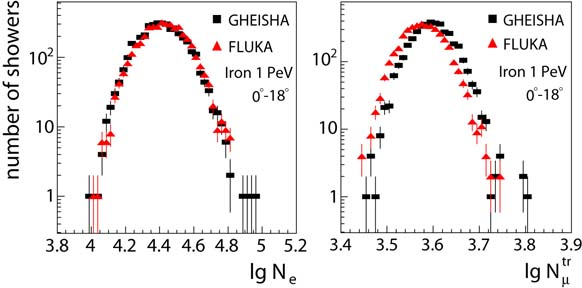}
\end{center}
\caption{Distribution of electron (left) and truncated muon (right) shower size for iron
induced showers of 1~PeV using GHEISHA and FLUKA as low energy interaction models and in both cases QGSJet as high energy interaction model.}
\label{qgsflugheiiron}
\end{figure}
\begin{figure}[h]
\begin{center}
\includegraphics*[width=0.55\textwidth]{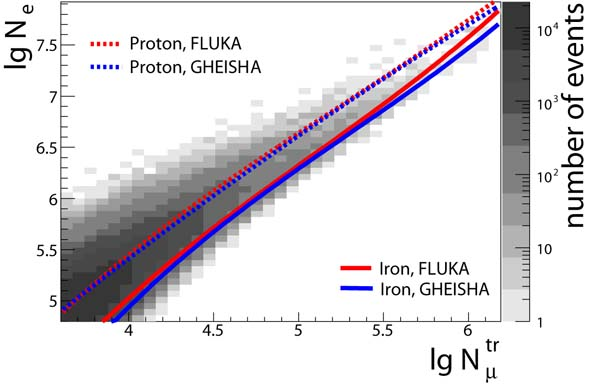}
\end{center}
\caption{Measured two-dimensional shower size spectrum ($-18^\circ$) together with lines of the most probable values for proton and iron induced showers for both simulations.}
\label{mostprob}
\end{figure}
GHEISHA has been widely used as low energy hadronic model in air shower simulations 
in the last decade. 

Though the high energy interaction models are believed to have a larger effect on 
the KASCADE observables, the influence of the low energy hadronic model ought to be 
investigated. In particular, the number of secondary muons at sea level could be affected
because these are decay products of low energy charged mesons. However, only at large 
distances from the core (where KASCADE is not sensitive), 
the contribution from low energy pions attains importance~\cite{Zab05}. 
It is known~\cite{Dres03,Hec03} that the GHEISHA model suffers from deficiencies in 
handling the reaction kinematics properly leading to, e.g., a notably flatter 
lateral distribution than the FLUKA model. 
In addition, GHEISHA produces too many pions as compared to accelerator data, 
while the FLUKA (version 2002.4)~\cite{FLU} package is in line with the measured data.

Therefore, in the more recent analysis of the KASCADE data GHEISHA is
replaced by the FLUKA model (preserving QGSJet version 2001 as high energy 
interaction model). Since FLUKA is known to describe the accelerator data to higher 
interaction energies, the change to the high energy interaction model is now at $200\,$GeV,
i.e.~FLUKA replaces in the intermediate energy range also the QGSJet model. 
 
When performing the simulations including detector response and the standard KASCADE
reconstruction, the differences between these two models are found to be rather 
small, as can be seen in Fig.~\ref{qgsflugheiiron}. 
As an example, the distributions of electron and truncated muon sizes for nearly 
vertical 1~PeV iron induced showers are displayed, one simulated
with the combination QGSJet/GHEISHA, the other with QGSJet/FLUKA. In case of FLUKA
simulations the shower size distribution is shifted by $\Delta\lg N_{e}\approx 0.01$
($\approx 2$\% for both, proton and iron induced showers)
towards larger electron numbers, and by $\Delta\lg N_{\mu}^{tr}\approx 0.02$ towards
($\approx 5$\% for iron induced EAS, but only $2$\% for proton induced EAS)
smaller muon numbers, i.e.~the differences are larger for the muon number than for the electron number.\footnote{Investigations of these kinds by using SIBYLL instead of
QGSJet as high energy interaction models lead for 1 PeV to a similar general behavior ($\approx 1$\% higher electron number in case of FLUKA and a smaller muon number), but interestingly for SIBYLL there is a $\approx 4$\% effect in muons for primary protons and only $\approx 2$\% for primary iron.}
For illustration in Fig.~\ref{mostprob} the most probable values of the simulated electron-muon number correlation for the two models and for different primaries are displayed on top of the data distribution. The differences are much smaller than compared to those observed for different high-energy interaction models (see Fig.~22 in ref.~\cite{Ant05}).
\begin{figure}[ht]
\includegraphics*[width=0.5\textwidth]{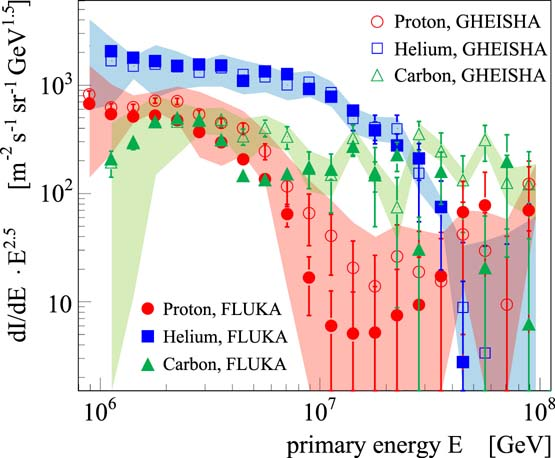}
\includegraphics*[width=0.5\textwidth]{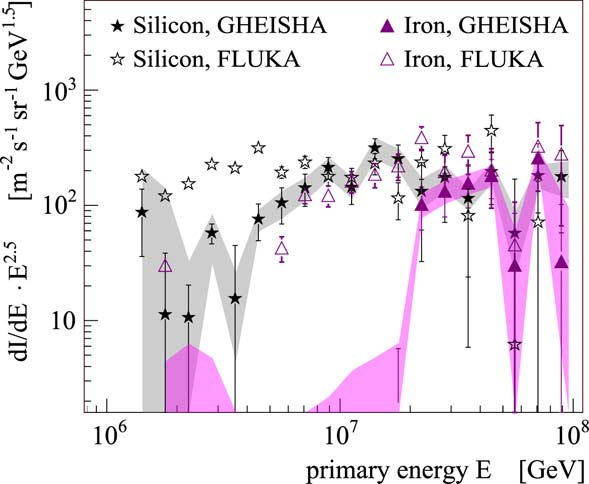}
\caption{Comparison between QGSJet/FLUKA based results and QGSJet/GHEISHA based results
for the energy spectra of H, He, and C (left) and of Si and Fe (right). 
Shaded bands correspond to estimates of the systematic uncertainties 
for the QGSJet/GHEISHA solutions.}
\label{gheiflu}
\end{figure}
\begin{figure}[h]
\includegraphics*[width=0.5\textwidth]{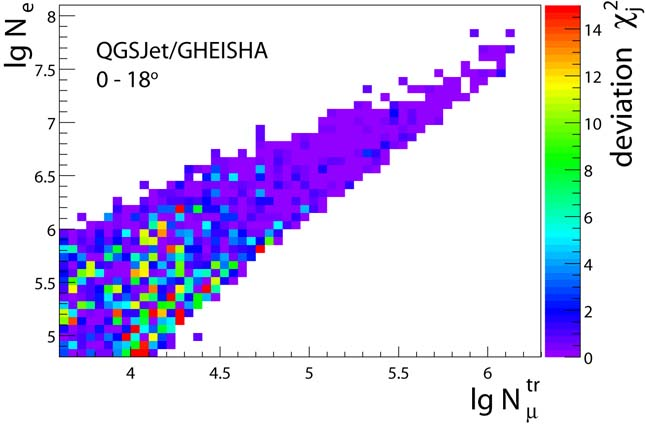}
\includegraphics*[width=0.5\textwidth]{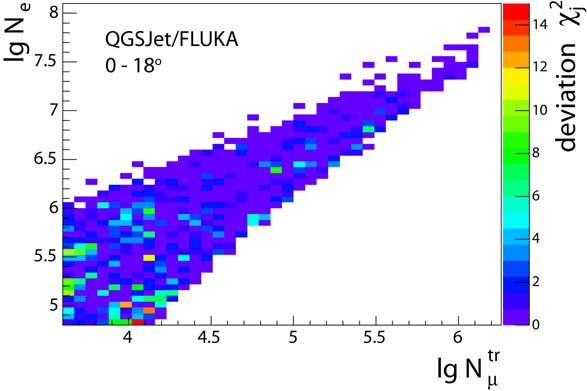}
\caption{Distribution of individual $\chi_j^2$ in the data range for the QGSJet solution. 
The low energy interaction model used is GHEISHA in the left panel (picture taken 
from~\cite{Ant05}) and FLUKA in the right panel.}
\label{chigheiflu}
\end{figure}
This behavior is expected in view of the above mentioned differences between 
GHEISHA and FLUKA: GHEISHA produces a higher pion multiplicity but flatter muon 
lateral distributions, or in other words, FLUKA predicts fewer muons in the range 
of $40\,$m - $200\,$m core distance. 

After carrying out the complete unfolding analysis, the results for the FLUKA case differ
only little from the GHEISHA case, as might have been expected from the small differences
in particle numbers. In Fig.~\ref{gheiflu}, left panel, the results for the energy
spectra of H, He, and C using GHEISHA and FLUKA are compared with each other. 
It is a specific characteristic of the QGSJet model that the analysis of 
the KASCADE data always results in a Helium dominated composition.  
Differences between the two solution sets are small, especially when compared to the 
systematic uncertainties imposed by the unfolding procedure~\cite{Ant05} 
(represented by the shaded bands in the figure, which are displayed 
for the QGSJet/GHEISHA solution only, but which are of the same order for all the discussed
solutions). 
This also holds for the corresponding spectra of Si and Fe, which are compared to each 
other in the right panel of Fig.~\ref{gheiflu}. Here, the influence of replacing 
the low energy interaction model on the result is larger than for the light elements. 
The relative abundance of the heavy elemental groups increases slightly when using FLUKA for the unfolding. This is understandable taking into account the shift of the simulated electron-muon numbers of primary iron towards the bulk of the data distribution (Fig.~\ref{mostprob}).
Finally, the results for the all-particle energy spectrum (which is the sum of 
the individual mass group spectra) are displayed in Fig.~\ref{allpartenspec} together 
with results discussed in the next section. 
No systematic difference in the all-particle spectrum could be observed by changing the low energy interaction model.

To test the quality of the solution the obtained mass spectra are folded forward
and compared to the measured data by a $\chi^2$-test~\cite{Ant05}. 
Noticeably, the $\chi^2_{dof}$ parameter improves considerably from $2.38$ to $1.34$ 
by going from GHEISHA to FLUKA. In Fig.~\ref{chigheiflu} the individual $\chi_j^2$ 
distributions for both cases are displayed, 
suggesting that FLUKA describes the correlation of muon to electron number in air showers better than GHEISHA.

To summarize, using the FLUKA model instead of the GHEISHA model seems to have no 
significant effect on the overall picture of the solution, but a slightly 
better description of the data is achieved.
This improvement in describing the KASCADE data is not surprising considering the fact that FLUKA has been tuned to provide a good description of recent accelerator data.

\section{Investigation of different KASCADE data sets}

In the analysis described so far, only EAS reaching the detector with zenith angles 
below 18$^{\circ}$ were included in the composition studies. 
The analysis of more inclined shower data can serve as consistency check.
Due to the limited reproduction of the measured observable correlations by the models, 
one cannot expect to obtain identical results for the energy spectra as compared 
to the vertical data set. Nevertheless, large differences between the solution sets 
for different zenith angle ranges could indicate a severe problem in either the 
simulation chain or the performed analysis technique. 
In particular, characteristics of the predictions of the attenuation of the 
electromagnetic and muonic shower components in the atmosphere can be tested. 
For this reason the analysis based on the QGSJet/FLUKA simulations were repeated for 
two additional data sets, the first one containing EAS with zenith angles ranging 
from 18$^{\circ}$ to 25.9$^{\circ}$, the second one from 25.9$^{\circ}$ to 32.3$^{\circ}$; 
thus covering the same acceptance on the sky. Data selection is identical to the one 
applied to the vertical data set, i.e.~the same runs in the same period and the same 
cuts (except for reconstructed zenith angle). 
The two-dimensional measured shower size spectra are included in Fig.~\ref{2dimspec}.

\begin{figure}[ht]
\begin{center}
\includegraphics*[width=0.49\textwidth]{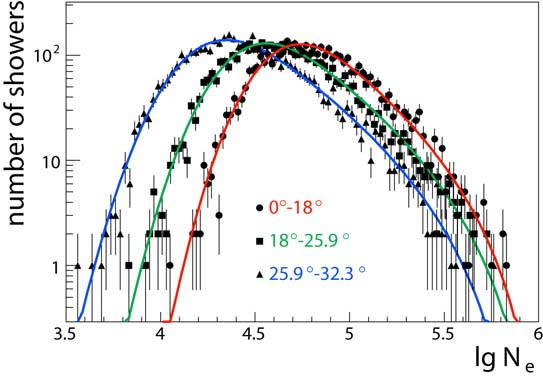}
\includegraphics*[width=0.49\textwidth]{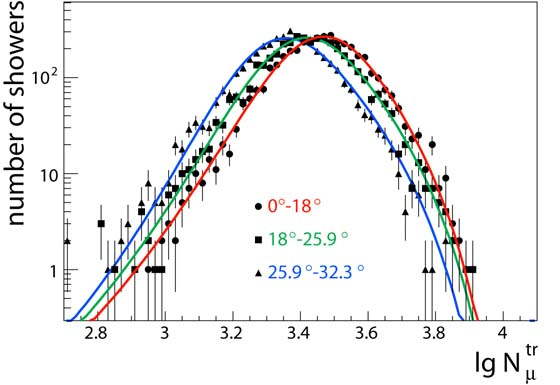}
\end{center}
\caption{Distribution of electron (left) and truncated muon (right) shower size for proton
induced showers of 1~PeV for different zenith angular ranges. The EAS are simulated 
with the QGSJet/FLUKA model combination. The lines show the parameterization of the 
distributions used in the analysis~\cite{Ant05}.}
\label{qgswinfluc}
\end{figure}

Again, first the shower fluctuations are compared to get an impression on the 
differences in the data sets. 
In Fig.~\ref{qgswinfluc}, as an example the distributions of electron and truncated
muon sizes for 1~PeV proton induced showers are displayed for all three angular ranges.
Going from nearly vertical to $\approx 30^\circ$ inclined showers, 
the shower size distribution is shifted by $\Delta\lg N_{e}\approx 0.5$ towards smaller electron numbers
(corresponding to an attenuation of the electron size relative to vertical EAS by $\approx 70$\%),
and by $\Delta\lg N_{\mu}^{tr}\approx 0.1$, 
(attenuation of muons relative to vertical EAS by $\approx 20$\%)
towards
smaller muon numbers. The widths of the distributions vary only slightly for 
the different angular ranges. Qualitatively, these differences are reflected in the 
two-dimensional shower size spectra shown in Fig.~\ref{2dimspec}, considering the 
coverage of the measured distribution in particle 
numbers.\footnote{The variation of the particle numbers with inclination is in first order independent of hadronic interaction models as it is due to attenuation in the atmosphere, which is handled by the CORSIKA code itself rather than by the interaction models.}
\begin{figure}[h]
\begin{center}
\includegraphics*[width=0.55\textwidth]{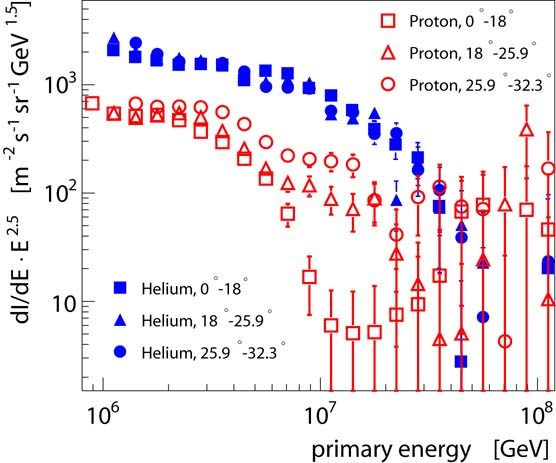}
\end{center}
\caption{Results for the energy spectra of primary H and He, based on the analysis of EAS data
originating from different zenith angle ranges (QGSJet/FLUKA). The display of the systematic 
uncertainties is omitted for reasons of clarity.}
\label{winres}
\end{figure}

By applying the full unfolding analysis for the resulting mass group spectra only small 
differences can be detected (Fig.~\ref{winres}). The display of the systematic error bands
is omitted in this figure for a better visibility. 
They are of the same order as those shown in Fig.~\ref{gheiflu} and therefore cover all 
three proton spectra, in particular at high energies. 
In the case of the proton spectra, systematic differences are present for 
energies above the proton knee. As can be seen, the knee in the proton spectrum gets less 
pronounced, i.e.~the change of index decreases with increasing zenith angle.
This may be related to the fact that shower fluctuations increase significantly for inclined showers and are larger also for lower primary masses.

For Helium, the most abundant group in all QGSJet based analyses~\cite{Ant05}, 
the spectra derived from the three data sets coincide within their 
statistical uncertainties. 
For the three heavier mass groups no significant difference within systematic and  
statistical accuracy could be found either. 
\begin{figure}[ht]
\includegraphics*[width=0.5\textwidth]{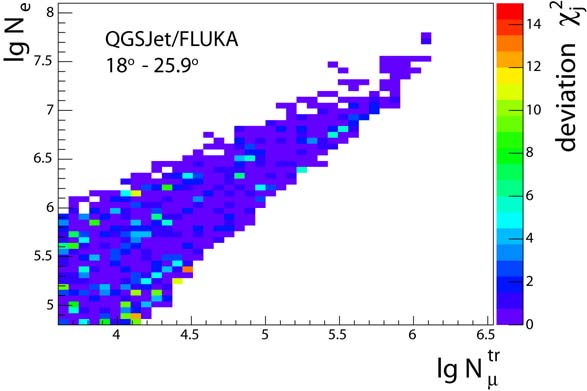}
\includegraphics*[width=0.5\textwidth]{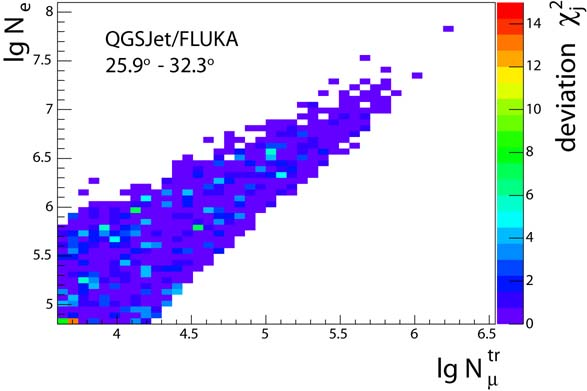}
\caption{Distribution of individual $\chi_j^2$ for the data of two different zenith 
angular ranges for the QGSJet/FLUKA solution.}
\label{chiangle}
\end{figure}

\begin{figure}[ht]
\begin{center}
\includegraphics*[width=0.55\textwidth]{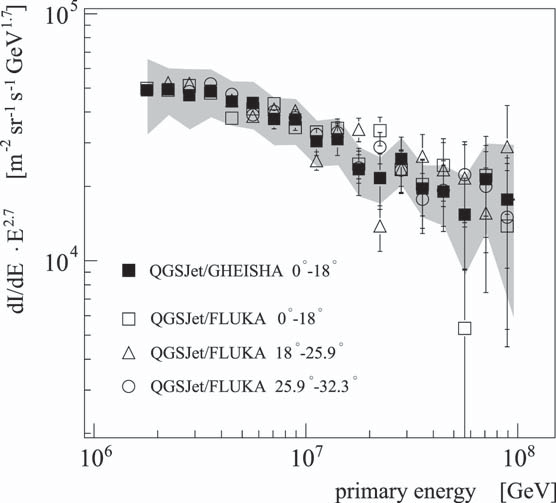}
\end{center}
\caption{Comparison between results for the all-particle energy spectrum, using GHEISHA
and FLUKA in the simulations. In the FLUKA case the data sets originate from
different zenith angle intervals. Error bars show the statistical uncertainty; 
the shaded band gives an estimate of the systematic uncertainties for the 
QGSJet/GHEISHA solution, which are of similar order for the other solutions.}
\label{allpartenspec}
\end{figure}

The observed systematic differences between the solution sets are small and can
be understood by the interplay of the dependence of shower sizes on energy and primary
particle type, increasing shower fluctuations with increasing zenith angle, and shifted
energy threshold (caused by the fixed data range in $N_{e}$ and $N_{\mu}^{tr}$) due to 
attenuation effects. The latter is also the reason for the increasing minimum 
primary energy of the results with increasing inclination of the incident showers.
The corresponding results for the all-particle energy spectrum are depicted in
Fig.~\ref{allpartenspec}. These coincide very well within their statistical uncertainties.

For completeness, in Fig.~\ref{chiangle} the $\chi_j^2$ distributions for the two 
inclined shower data sets are shown. The $\chi^2_{dof}$ parameter, 
which is $1.34$ for vertical showers improves further to $1.17$ for the range 
$18^\circ - 25.9^\circ$ and to $0.92$ for the third range.  
This could be expected following the arguments above, as the QGSJet model was 
shown to have problems in describing the data at lower energies. 
When going to larger inclinations, keeping the cuts on measured particle numbers, 
results in a higher energy threshold. 

\section{Summary}

The analysis in terms of energy spectra of individual mass groups as described 
in~\cite{Ant05} has been applied to KASCADE data based on a different low energy 
hadronic interaction model and to data sets in different intervals of the zenith angle.
 
Using the FLUKA model instead of the GHEISHA model has no significant effect on the 
overall picture of the solution, but a slightly better description of the data is achieved,
i.e.~i.e.~the correlation of electron to muon number in the simulations with FLUKA is in better agreement with the data which results in a lower $\chi^2_{dof}$ distribution of the solution.
The results of the unfolding analyses of the two-dimensional shower size
spectrum for different zenith angular ranges show no strong or unexplainable systematic
differences. Thus, the results give no hint to any severe problem in the
simulation or the analysis, and reaffirm the conclusions~\cite{Ant05} drawn from 
the analysis of the nearly vertical shower set:
The knee is observed at an energy around $\approx 5\,$PeV with a change of the index 
$\Delta \gamma \approx 0.4\,$. Considering the results of the mass group spectra, 
in all analyses an appearance of knee-like features in the spectra of the light 
elements is ascertained. 
In all solutions the positions of the knees in these spectra is shifted to higher energy
with increasing element number.

By applying the analysis to different data sets and based on different interaction models,
it has been demonstrated that unfolding methods are capable to reconstruct energy spectra 
of individual mass groups from air shower data, in addition to the all-particle spectrum. 
But still, the limiting factor of the analysis are the properties of the 
hadronic interaction models used and not the quality or the understanding of the 
KASCADE data. 
Furthermore, the procedure of the KASCADE data analysis, and in future also the analysis of KASCADE-Grande data measuring higher primary energies and muons at larger distances~\cite{Zab05}, gives valuable hints  
for the improvement of hadronic interaction models.
The data can be confidently used when improved interaction models, based on more and extended 
accelerator experiments, become available.

\begin{ack}
This work is supported by the BMBF of Germany, the Ministry of Science and Higher Education
of Poland, and the Romanian Ministry of Education and Research 
(grant CEEX 05-D11-79/2005).
\end{ack}

\end{document}